\documentclass[conference]{IEEEtran}
\IEEEoverridecommandlockouts

\usepackage{algorithm,algorithmic,amssymb,amsthm,bbm,cite,color,enumitem,graphicx,microtype,setspace,tabularray,url}
\usepackage[tbtags]{amsmath}
\usepackage[USenglish]{babel}
\usepackage[bookmarks=false]{hyperref}
\usepackage[utf8]{inputenc}
\usepackage[T1]{fontenc}

\usepackage[labelsep=period,font=small]{caption}
\usepackage[nolist]{acronym}
\usepackage{cite}
\usepackage{placeins}
\usepackage{tikz,pgfplots}
\usetikzlibrary{arrows, arrows.meta, backgrounds, calc, intersections, decorations.markings, decorations.pathreplacing, positioning, shapes}

\usetikzlibrary{external}
\pgfplotsset{compat=1.17,
legend image code/.code={
\draw[mark repeat=2,mark phase=2]
plot coordinates {
(0cm,0cm)
(0.15cm,0cm)        
(0.3cm,0cm)         
};%
}
}

\usepackage{titlesec}
\let\subparagraph\relax
\titlespacing{\section}{0pt}{6pt plus 2pt minus 1pt}{4pt plus 1pt minus 1pt} 
\titlespacing{\subsection}{0pt}{4pt plus 2pt minus 1pt}{2pt plus 1pt minus 1pt} 
\newcommand{\h}{\mathbf{h}}

\renewcommand{\v}{\mathbf{v}}

\newcommand{\z}{\mathbf{z}}

\newcommand{\0}{\mathbf{0}}

\newcommand{\setC}{\mathcal{C}}

\newcommand{\setN}{\mathcal{N}}

\newcommand{\Real}{\mbox{$\mathbb{R}$}}
\newcommand{\Compl}{\mbox{$\mathbb{C}$}}

\newcommand{\diff}{\mathop{}\!\mathrm{d}}
\newcommand{\Exp}{\mathbb{E}}
\newcommand{\herm}{\mathrm{H}}

\mathchardef\mhyphen="2D

\def \am {$\alpha$-$\mu$}
\def \km {$\kappa$-$\mu$}
\def \hm {$\eta$-$\mu$}
\def \Ehm {Extended $\eta$-$\mu$}
\def \ahkm {$\alpha$-$\eta$-$\kappa$-$\mu$}

\definecolor{myred}{HTML}{F05039}
\definecolor{myorange}{HTML}{EEBAB4}
\definecolor{myblue}{HTML}{3D65A5}
\definecolor{mygreen}{HTML}{539054}
\definecolor{myyellow}{HTML}{F0E442}

\let\originalleft\left
\let\originalright\right
\renewcommand{\left}{\mathopen{}\mathclose\bgroup\originalleft}
\renewcommand{\right}{\aftergroup\egroup\originalright}

\begin{acronym}

\acro{3GPP}{the 3rd Generation Partnership Project}
\acro{5G}{fifth-generation}
\acro{6G}{sixth-generation}
\acro{AWGN}{additive white Gaussian noise}
\acro{BEP}{bit error probability}
\acro{BER}{bit error rate}
\acro{BFSK}{binary frequency-shift keying}
\acro{BPSK}{binary phase-shift keying}
\acro{BS}{base station}
\acro{CCDF}{complementary cumulative distribution function}
\acro{CDF}{cumulative distribution function}
\acro{CP}{coverage probability}
\acro{CSI}{channel state information}
\acro{EGC}{equal gain combining}
\acro{FTR}{fluctuating two-ray}
\acro{i.i.d.}{independent and identically distributed}
\acro{LoS}{line-of-sight}
\acro{$M$-PSK}{$M$-ary phase-shift keying}
\acro{$M$-QAM}{$M$-ary quadrature amplitude modulation}
\acro{MGF}{moment-generating function}
\acro{mmWave}{millimeter-wave}
\acro{MRC}{maximum ratio combining}
\acro{MRT}{maximum ratio transmission}
\acro{NLoS}{non-line-of-sight}
\acro{OP}{outage probability}
\acro{PDF}{probability density function}
\acro{RF}{radio frequency}
\acro{RIS}{reflective intelligent surface}
\acro{RV}{random variable}
\acro{SEP}{symbol error probability}
\acro{SER}{symbol error probability}
\acro{SISO}{single-input single-output}
\acro{SNR}{signal-to-noise ratio}
\acro{sub-THz}{sub-terahertz}

 \end{acronym}

\title{A New Framework for the Sum of Squared $\kappa$-$\mu$ RVs with Application to Sub-THz Systems}
\author{
\IEEEauthorblockN{Gustavo Rodrigues de Lima Tejerina and Italo~Atzeni}
\IEEEauthorblockA{Centre for Wireless Communications, University of Oulu, Finland \\
E-mail: \{gustavo.tejerina, italo.atzeni\}@oulu.fi}
\thanks{This work was supported by the Research Council of Finland (336449 Profi6, 348396 HIGH-6G, and 369116 6G~Flagship) and by the European Commission (101095759 Hexa-X-II).}}

\begin{document}

\maketitle

\begin{abstract}
In this paper, we adopt the \km{} model to characterize the propagation in the sub-THz band. We develop a new exact representation of the sum of squared independent and identically distributed \km{} random variables, which can be used to express the power of the received signal in multi-antenna systems. Unlike existing ones, the proposed analytical framework is remarkably tractable and computationally efficient, and thus can be conveniently employed to analyze systems~with~massive~antenna~arrays. We derive novel expressions for the probability density function and cumulative distribution function, analyze their convergence and truncation error, and discuss the computational complexity and the implementation aspects. Moreover, we derive expressions for the coverage probability and bit error probability for coherent binary modulations. Lastly, we evaluate the performance of an uplink sub-THz system where a single-antenna user is served by a base station employing maximum ratio combining.
\end{abstract}


\vspace{-0.5mm}

\section{Introduction} \label{sec:Intro}

6G-and-beyond wireless systems are envisioned to explore the sub-THz spectrum, ranging from 100 to 300~GHz, to seek extreme bandwidths and accommodate an unprecedented number of users with high data rates~\cite{Raj20}. In general, raising the carrier frequency has a two-fold impact on the propagation~\cite{Han22}. First, the penetration loss and roughness of the materials with respect to the wavelength increase: consequently, the impact of the \ac{NLoS} paths tends to diminish and the \ac{LoS} component becomes dominant. Second, the free-space path loss (for isotropic antennas) increases: thus, more antennas are needed to create physically large arrays that are capable of properly focusing the signal power. 

Statistical channel characterization in the sub-6~GHz range typically employs the classical Rayleigh, Rice, and Nakagami-$m$ models.
However, such models fail to properly capture the intricacies of the propagation at high frequencies, which motivates investigating more generalized ones such as the (Extended) \hm{}, \km, \am{}, \ahkm{}, and \ac{FTR} models. For instance,~\cite{Mar21,Pap21,Du22} compared both classical and generalized fading models to measured data at 26, 28, 39, and 143.1~GHz in \ac{LoS} and \ac{NLoS} scenarios, showing a better fit for the generalized models. Other works investigated generalized fading models applied to specific scenarios including, e.g., multiple antennas~\cite{Jos22}, reflective surfaces~\cite{Le24}, dual-hop communications~\cite{Li22}, and random atmospheric absorption~\cite{Bha24}.

In this paper, we adopt the \km{} model~\cite{Yac07} to characterize the propagation in the sub-THz band. Indeed, the \km{} model provides remarkable flexibility to represent scenarios with a dominant \ac{LoS} component, which well fit the propagation at sub-THz frequencies. Moreover, it encompasses classical models such as Nakagami-$m$, Rice, and Rayleigh. In multi-antenna systems with \ac{MRT} at the transmitter or \ac{MRC} at the receiver, the power of the received signal can be expressed as a sum of squared \acp{RV}. In this regard,~\cite{Mil08} proposed expressions for the \ac{PDF} and \ac{CDF} of the sum of squared \ac{i.i.d.} \km{} \acp{RV} employing special functions such as the modified Bessel function and Marcum Q-function. Additionally, evaluating the corresponding performance metrics either involves Horn functions~\cite{Dix23}, which are not readily available in common mathematical packages (such as Mathematica and SciPy), or requires cumbersome numerical integration. Hence, existing analytical frameworks become unsuitable to handle massive numbers of antennas due to their inherent computational~complexity.

To address this issue, we extend the framework in~\cite{Alm23a} and develop a new exact representation of the sum of squared \ac{i.i.d.} \km{} \acp{RV}. The resulting framework is remarkably tractable and computationally efficient, and thus can be conveniently employed to study sub-THz systems with massive antenna arrays.\footnote{Further results for the \km{} model, along with an entirely new analysis using the \Ehm{} model applied to FR3 systems, can be found in the longer version of this paper~\cite{Tej25}.} We derive novel expressions for the \ac{PDF} and \ac{CDF}, analyze their convergence and truncation error, and discuss the computational complexity and implementations aspects. Then, we derive expressions for the coverage probability and \ac{BEP} for coherent binary modulations, along with their asymptotic expressions at high \ac{SNR}. Lastly, we evaluate the performance of an uplink sub-THz system under the \km{} model, where a single-antenna user is served by a multi-antenna \ac{BS} employing \ac{MRC}, with perfect and imperfect \ac{CSI}. For example, we observe that a $4 \times$ increase in the number of antennas allows the carrier frequency to be increased by about 60\% while maintaining the same \ac{BEP}. Moreover, even in \ac{LoS}-dominated scenarios, the \ac{NLoS} components continue to have a noticeable impact on the system's performance.

\smallskip

\textit{\textbf{Notation.}} $f_X(x)$, $F_X(x)$, and $\mathcal{M}_X(s)$ denote the \ac{PDF}, \ac{CDF}, and \ac{MGF}, respectively, of \ac{RV} $X$. $\text{erfc}(\cdot)$ is the complementary error function~\cite[Eq.~(06.27.07.0001.01)]{WolframResearch}, $\Gamma(\cdot)$ is the Gamma function~\cite[Eq.~(06.05.02.0001.01)]{WolframResearch} and $\Gamma(\cdot,\cdot)$ is the incomplete Gamma function~\cite[Eq.~(06.06.02.0001.01)]{WolframResearch}. $I_\nu(\cdot)$ denotes the modified Bessel function of the first kind and $\nu$-th order~\cite[Eq.~(03.02.02.0001.01)]{WolframResearch}. $_p F_q (a_1,\ldots,a_p;b_1,\ldots,b_q;z)$, for $|z|<1$ if $q=p-1$ or $\forall z \in \Real$ if $q \geq p$, is the generalized hypergeometric function~\cite[Eq.~(07.31.02.0001.01)]{WolframResearch}. Lastly, $\sim$ means asymptotically equivalent.

\section{New Framework for the Sum of Squared \km{} \acp{RV}}\label{sec:newrep}

Considering an $N$-antenna \ac{BS} serving a single-antenna user, the sum of squared \acp{RV} can be used to express the instantaneous \ac{SNR} in the downlink when the \ac{BS} transmits data using \ac{MRT} or in the uplink when the \ac{BS} receives data using \ac{MRC}. In this paper, we focus on the latter as our motivating scenario and for our performance evaluation, although the proposed framework also applies to the other setting~\cite{Tej25}. In this context, the received signal after combining is given by
\begin{align}
y = \v^{\herm} (\sqrt{P_{\textrm{t}}}\h x + \z) \in \Compl,
\end{align}
where $\v \in \Compl^{N}$ is the combining vector, $P_{\textrm{t}}$ is the transmit power, $\h = [h_{1}, \ldots, h_{N}] \in \Compl^{N}$ is the channel vector, $x \in \Compl$ is the transmitted data symbol (with $\Exp\big[ |x|^{2} \big] = P_{\textrm{t}}$), and $\z \in \Compl^{N}$ is the noise term with \ac{i.i.d.} $\setC \setN(0, \sigma^{2})$ entries. With perfect \ac{CSI}, the \ac{MRC} combining vector is given by $\v = \h$ and the corresponding \ac{SNR} at the \ac{BS} is
\begin{align} \label{eq:perfectSNR}
\textrm{SNR}_{\textrm{P}} = \frac{P_{\textrm{t}} |\v^{\herm} \h|^{2}}{\sigma^{2} \| \v \|^{2}} = \frac{P_{\textrm{t}}}{\sigma^{2}} \sum_{n=1}^{N} |h_{n}|^{2}.
\end{align}
Alternatively, assuming a simplified model for imperfect \ac{CSI}, the \ac{MRC} combining vector is given by $\v = \hat{\h}$, where $\hat{\h} = \sqrt{1-\alpha^{2}} \h + \alpha\tilde{\h} \in \Compl^N$ denotes the estimated channel, $\tilde{\h} \in \Compl^{N}$ is the channel estimation error (independent of $\h$) with \ac{i.i.d.} $\setC \setN \big(\0, \frac{1}{N} \| \h \|^{2}\big)$ entries, and $\alpha \in [0,1]$ is the estimation accuracy (with $\alpha=0$ representing perfect \ac{CSI}). When $N \gg 1$, the corresponding \ac{SNR} at the \ac{BS} can be approximated as $\textrm{SNR}_{\textrm{I}} \simeq (1-\alpha^{2}) \textrm{SNR}_{\textrm{P}}$, with $\textrm{SNR}_{\textrm{P}}$ defined in \eqref{eq:perfectSNR}.

Motivated by the above system model, we propose a new exact representation of the sum of squared \ac{i.i.d.} \km{} \acp{RV}. The proposed framework is based on the recursive solution provided in~\cite{Alm23a} and results in much simpler expressions for the \ac{PDF} and \ac{CDF} (i.e., given in terms of simple Gamma functions) compared with~\cite{Mil08,Dix23}. Hence, the resulting performance metrics, such as the \ac{BEP}, also inherit the framework's simplicity. Furthermore, we analyze the convergence, truncation error, and computational complexity of these new expressions. The proposed framework has proven to return fast and precise solutions even with very large numbers of antennas.

\subsection{\ac{PDF} and \ac{CDF}} \label{sec:newrep_km}

The \km{} model is a generalized fading model used to represent scenarios with a dominant \ac{LoS} component~\cite{Yac07}. In this model, $\kappa > 0$ denotes the power ratio between the \ac{LoS} and \ac{NLoS} (i.e., scattered) components, whereas $\mu > 0$ indicates the number of multipath clusters. For the \km{} distribution, the \ac{PDF} of the squared envelope of the $n$-th \ac{RV}, denoted by $W_{n} = \frac{P_{\textrm{t}}}{\sigma^{2}} |h_{n}|^2$, is defined as~\cite{Yac07}
\begin{align} \label{eq:kappa:pdfsnr}
    f_{W_n}(w_n) = \frac{ (\kappa +1)^{\frac{\mu +1}{2}} w_n^{\frac{\mu-1}{2}} I_{\mu -1}\big(2\sqrt{\kappa \mu K \frac{w_{n}}{\hat{w}_{n}}}\big)}{\mu^{-1}\kappa ^{\frac{\mu -1}{2}} \hat{w}_n ^{\frac{\mu +1}{2}} \exp \big(\kappa  \mu + \frac{K w_{n}}{\hat{w}_{n} }\big)},
\end{align}
with $w_n > 0$, $K = (1+\kappa)\mu$, and $\hat{w}_{n} = \mathbb{E}[W_n]$. For the new representation of the sum of $N$ squared \ac{i.i.d.} \km{} \acp{RV}, we follow the standard \ac{MGF}-based approach to obtain the \ac{PDF}: first, we derive the \ac{MGF} of the squared \ac{RV} and then we express the sum as the product of $N$ \acp{MGF}.

For the first step, we begin by obtaining the \ac{MGF} of \eqref{eq:kappa:pdfsnr} after applying the Laplace transform, i.e., $ \mathcal{M}_{X}(s) = \int_0^\infty f_X(x)\exp\left( -s x \right)\diff x$, as
\begin{align}\label{eq:kappa:mgf-first}
        \mathcal{M}_{W_n}(s) = & \ \frac{\mu  (\kappa +1)^{\frac{\mu +1}{2}}}{\kappa ^{\frac{\mu -1}{2}} \exp (\kappa  \mu ) \hat{w}_{n} ^{\frac{\mu +1}{2}}}\int_0^{\infty} \frac{w_n^{\frac{\mu -1}{2}}}{\exp(s w_n)} \nonumber\\
    & \times \frac{I_{\mu -1}\big(2 \sqrt{ \kappa \mu K \frac{w_n}{\hat{w}_n }}\big)}{\exp\big(\frac{K w_n}{\hat{w}_n }\big)}  \diff w_n .
\end{align}
Following the same rationale of~\cite{Alm23a}, we rewrite the exponential and modified Bessel functions according to their contour integral representations~\cite[Eq.~(03.02.07.0009.01), (01.03.07.0001.01)]{WolframResearch}, which yields
\begin{align}\label{eq:kappa:mgf-contour1}
        \mathcal{M}_{W_n}(s) = \ &\frac{\jmath^{-1-\mu}\mu (\kappa +1)^{\frac{\mu +1}{2}}}{4\pi^{2}  \kappa ^{\frac{\mu -1}{2}} \hat{w}_n ^{\frac{\mu +1}{2}} \exp (\kappa  \mu )}\! \int _0^{\infty }  \!\oint_{\mathcal{L}_t} \!\oint_{\mathcal{L}_v} \frac{w_n^{\frac{\mu -1}{2}} \Gamma (t)}{(s w_n)^{t}} \nonumber\\
        &\times  \frac{ \Gamma \big(v+\frac{\mu -1}{2}\big) \big(\!- \frac{ \kappa \mu K w_n}{\hat{w}_n }\big)^{-v}}{\Gamma \big(\frac{\mu -2 v+1}{2}\big) \exp \big( \frac{K w_n}{\hat{w}_n }\big) } \diff v \diff t \diff w_n,
\end{align}
where $\mathcal{L}_t$ and $\mathcal{L}_v$ are the complex contours of $t$ and $v$, respectively. The integral operators in \eqref{eq:kappa:mgf-contour1} are then reordered to solve the first one in terms of $w_n$, resulting in
\begin{align}
    \mathcal{M}_{W_n}(s) = \ &\frac{\jmath^{-1-\mu} (\kappa  \mu )^{\frac{1-\mu }{2}} }{4\pi^{2}\exp (\kappa  \mu )} \!\oint_{\mathcal{L}_t^{*}} \!\oint_{\mathcal{L}_v^{*}} \frac{ \Gamma (t) \Gamma \big(\frac{\mu -1}{2}+v\big)}{\Gamma \big(\frac{\mu + 1}{2}-v\big) (-\kappa  \mu )^{v}} \nonumber\\
        & \times \Gamma \bigg(\frac{\mu}{2} - t- v+\frac{1}{2}\bigg) \bigg(\frac{K }{s \hat{w}_n }\bigg)^t \diff v \diff t, \label{eq:kappa:mgf-contour2}
\end{align}
where $\mathcal{L}_t^{*}$ and $\mathcal{L}_v^{*}$ are the new complex contours of $t$ and $v$, respectively, after the integration over $w_n$. At this stage, \eqref{eq:kappa:mgf-contour2} has new essential singularities located at $t \in \{-m, (-2 l+\mu +2 m+1)/2\}$ and at $v \in \{(-2 l-\mu +1)/2,(-2 l+\mu +2 m+1)/2\}$. As in~\cite{Alm23a}, by identifying the singularities, the new complex contours are defined to guarantee the convergence of \eqref{eq:kappa:mgf-contour2} and circumvent duplicate poles. With this in mind, both contour integrals can be solved by applying the residue theorem~\cite[Eq.~(16.3.6)]{Kre10} to \eqref{eq:kappa:mgf-contour2} for the poles $t\to 1/2 (\mu +2 m-2 v+1)$ and $v\to 1/2 (-2 l-\mu +1)$, leading to
\begin{align}\label{eq:kappa:mgf-sum1}
       \mathcal{M}_{W_n}(s) =\ & \bigg(\frac{K}{\exp(\kappa) s \hat{w}_n }\bigg)^{\mu } \sum_{m=0}^{\infty} \sum_{l=0}^{\infty} \frac{\Gamma (l+m+\mu )}{m! l! \Gamma (l+\mu )} \nonumber\\
       &\times \left(\frac{\kappa  \mu K}{s \hat{w}_n }\right)^l \big(-\frac{K }{s \hat{w}_n }\big)^m. 
\end{align}
Finally, \eqref{eq:kappa:mgf-sum1} is rearranged in terms of Cauchy product as

\clearpage

$ $ \vspace{-9mm}

\begin{align}\label{eq:kappa:mgf-sum2}
 \mathcal{M}_{W_n}(s) = \ & \bigg(\frac{K }{\exp(\kappa) s \hat{w}_n }\bigg)^{\mu } \sum_{m=0}^{\infty}\bigg(\frac{1}{s \hat{w}_n} \bigg)^{m}\sum_{l=0}^{m} \bigg(\!-\frac{1}{\kappa  \mu }\bigg)^l \nonumber\\
        &\times \frac{(\kappa  \mu K)^m \Gamma (m+\mu )}{l! (m-l)! \Gamma (m-l+\mu )}.
\end{align}
For the second step, the \ac{MGF} of the sum of $N$ \ac{i.i.d.} \acp{RV} $W_{n}$, denoted by $W = \sum_{n=1}^{N} W_{n} = \textrm{SNR}_{\textrm{P}}$, is obtained as the product of their respective \acp{MGF}, i.e., $\mathcal{M}_{W} (s) = \prod_{n=1}^{N} \mathcal{M}_{W_n} (s)$. From \eqref{eq:kappa:mgf-sum2} and~\cite[Eq.~(0.314)]{Gra07}, this product is given by
\begin{align} \label{eq:kappa:mgf}
        \mathcal{M}_W (s) = \left(\frac{(\kappa +1) \mu }{\exp(\kappa) }\right)^{N \mu } \sum_{m=0}^{\infty}\left(\frac{1}{s \hat{w}}\right)^{N \mu + m} k_m,
\end{align}
where we have replaced $\hat{w}_{n} = \hat{w}$, $\forall n$ due to the \ac{i.i.d.} property and with
\begin{subequations} \label{eq:km}
\begin{align}
    k_0 = \ & 1, \label{eq:km0} \\
    k_m = \ & \frac{1}{m} \sum_{i=1}^{m}(i-m+N i) \Gamma (i+\mu ) k_{m-i} \nonumber \\
    & \times \sum_{l=0}^i \frac{\left(-\kappa  \mu \right)^{-l} \left(\kappa  (\kappa +1) \mu ^2\right)^i}{l! (i-l)! \Gamma (-l+i+\mu )}, \, m\geq 1. \label{eq:km1}
\end{align}
\end{subequations}
Finally, the inverse Laplace transform is applied to \eqref{eq:kappa:mgf} by using~\cite[Eq.~(17.13.3)]{Gra07} and the \ac{PDF} is obtained as
\begin{align} \label{eq:pdf_snr}
        f_{W}(w) = \left(\frac{(\kappa +1) \mu }{\exp(\kappa) }\right)^{N \mu } \sum_{m=0}^{\infty}\frac{ w^{N \mu + m-1} k_m}{ \hat{w}^{N \mu+ m} \Gamma (N \mu + m )},
\end{align}
with $k_m$ defined in \eqref{eq:km}. The \ac{CDF} readily follows from applying $F_X(x) = \int_0^{x} f_X(x)\diff x$, which yields
\begin{align} \label{eq:cdf_snr}
        F_{W}(w) = \left(\frac{(\kappa +1) \mu }{\exp(\kappa) }\right)^{N \mu } \sum_{m=0}^{\infty}\frac{\big(\frac{w}{\hat{w}}\big)^{N \mu + m} k_m}{\Gamma (N \mu + m + 1)}.
\end{align}

An alternative \ac{MGF} representation can be obtained from \eqref{eq:kappa:mgf-sum1} by simplifying its $m$-indexed summation and reducing it to a power function, resulting in 
\begin{align} \label{eq:km:mgf-pb2}
        \mathcal{M}_{W}(s)&=  \sum_{m=0}^{\infty} \frac{\tilde{k}_m}{\exp(N \kappa \mu)} \bigg(\frac{  K }{ K + s \hat{w} }\bigg)^{N\mu +m} ,
\end{align}
with (cf. \eqref{eq:km})
\begin{subequations} \label{eq:km:pb-km}
\begin{align}
    \tilde{k}_0 = \ & 1, \label{eq:km:pb-km-0} \\
    \tilde{k}_m = \ & \frac{1}{m} \sum _{i=1}^m \frac{(N i +i-m)(\kappa \mu )^i \tilde{k}_{m-i}}{\Gamma(i+1)}, \, m\geq 1. \label{eq:km:pb-km-m}
\end{align}
\end{subequations}
From \eqref{eq:km:mgf-pb2}, the PDF is obtained as (cf. \eqref{eq:pdf_snr})
\begin{align} \label{eq:kappa:pdf_snr2}
        f_{W}(w) = \frac{\exp\big(\!-\frac{K w}{\hat{w}}\big)}{\exp(N\kappa\mu) } \sum_{m=0}^{\infty}\frac{ \big(\frac{K w}{\hat{w}}\big)^{N \mu + m} \tilde{k}_m}{ w \Gamma (N \mu + m )},
\end{align}
whereas the CDF is given by (cf. \eqref{eq:cdf_snr})
\begin{align} \label{eq:kappa:cdf_snr2}
        F_{W}(w) = \sum_{m=0}^{\infty}\Bigg(1-\frac{ \Gamma \big(N \mu + m,\frac{ K w}{\hat{w} }\big)}{ \Gamma (N \mu + m )}\Bigg)\frac{\tilde{k}_m}{\exp(N\kappa\mu)}.
\end{align}

In Section~\ref{sec:metrics}, we will use the \acp{PDF} in \eqref{eq:pdf_snr} and \eqref{eq:kappa:pdf_snr2} to derive two distinct expressions for the \ac{BEP}: one with lower computational complexity but quite restrictive convergence conditions, and another with slightly higher computational complexity and no restrictions on the convergence.

\subsection{Convergence and Truncation Analysis} \label{sec:trunc}

We now analyze the infinite summations in \eqref{eq:pdf_snr}--\eqref{eq:cdf_snr} in terms of absolute convergence, which occurs when the sum of the absolute values of the individual terms converges~\cite[Eq.~(0.21)]{Gra07}. Hence, we obtain an upper bound on the summations in \eqref{eq:pdf_snr}--\eqref{eq:cdf_snr}, as done in~\cite{Alm23a}. Based on this, we derive the truncation error associated with the \ac{PDF} and \ac{CDF}, which provides useful insights into the number of terms needed to numerically compute these functions with the desired accuracy.

For the summations in \eqref{eq:pdf_snr}--\eqref{eq:cdf_snr}, we need ensure that
\begin{align} \label{eq:kappa:ub}
   \mathcal{C} = \sum_{m=0}^{\infty}\frac{w^{N \mu + m-1+\zeta} |k_m|}{\hat{w} ^{N \mu + m}\Gamma (N \mu + m +\zeta)}  < \infty,
\end{align}
where $\zeta = 0$ and $\zeta = 1$ correspond to \eqref{eq:pdf_snr} and \eqref{eq:cdf_snr}, respectively. To this end, we first upper bound $\left|k_m\right|$ for $m\geq 1$ as \vspace{-0.5mm}
\begin{align} \label{eq:kappa:up1}
    \left| k_m\right| < \ & \sum_{i=1}^{m} \sum_{l=0}^i \frac{(i-m+N i) \Gamma (i+\mu )  K^i |k_{m-i}|}{m l! (i-l)! \Gamma (i-l+\mu ) \left(\kappa  \mu \right)^{l-i}}.
\end{align}
As $1/\Gamma(i-l+\mu) \leq 1/\Gamma(\mu)$ for $l\in [1,i]$ and $1/\Gamma(\mu) < 8/7$ for any choice of $\mu$, we can upper bound the $l$-indexed summation of \eqref{eq:kappa:up1} by a power function~\cite[Eq.~(1.111)]{Gra07}, obtaining
\begin{align} \label{eq:kappa:up2}
    | k_m| <\ & \frac{8}{7m} \sum _{i=1}^m \frac{( i -m+ N i) \Gamma (i+\mu ) |c_{m-i}|}{\Gamma(i+1) \tilde{K}^{-i}},
\end{align}
with $\tilde{K} = K (\kappa\mu+1)$. Considering the last term of the summation (corresponding to $i = m$) and the fact that $1/\Gamma(m+1) < 1/\Gamma(m)$, we obtain the simpler upper bound 
\begin{align} \label{eq:kappa:up3}
    | k_m| < \ \frac{}{} \frac{ 8 N\Gamma (\mu +m )}{ 7 \Gamma(m) \tilde{K}^{-m}}.
\end{align}
Now, we extend \eqref{eq:kappa:up3} to all $m$ by substituting the right-hand side of \eqref{eq:kappa:up3} into \eqref{eq:kappa:ub} and performing the index change $m = i + 1$. After some mathematical manipulations, we obtain
\begin{align}
    \mathcal{C} < \ & 
    \frac{w^{N \mu-1+\zeta}}{\hat{w} ^{N \mu} \Gamma (N \mu +\zeta )} \bigg( 1 +\frac{8 N  \Gamma (\mu +1) \Gamma (N \mu +\zeta) w}{7 \Gamma (N \mu +1 + \zeta) \hat{w}} \nonumber \\
    & \times \tilde{K} \, _1F_1\bigg(\mu +1;N \mu+1+\zeta;\frac{\tilde{K} w}{ \hat{w}}\bigg)\bigg). \label{eq:kappa:ub3}
\end{align}
Finally, the absolute convergence of \eqref{eq:kappa:ub} and, consequently, the convergence of \eqref{eq:pdf_snr}--\eqref{eq:cdf_snr}, follows from observing that the Kummer confluent hypergeometric function $_1F_1$ in \eqref{eq:kappa:ub3} is finite for any choice of parameters in our framework.

Moreover, we derive an upper bound on the truncation error when applied to \eqref{eq:pdf_snr}--\eqref{eq:cdf_snr}. We begin by defining the truncation error as the sum of the truncated terms, i.e.,
\begin{align} \label{eq:kappa:pdferror}
        \mathcal{E}(\epsilon) = \bigg(\frac{K}{\exp(\kappa)} \bigg)^{N \mu} \sum _{m=\epsilon}^{\infty } \frac{\big(\frac{w}{\hat{w}}\big)^{N \mu+m} k_m}{w ^{1-\zeta } \Gamma (N \mu + m +\zeta )},
\end{align}
where $\epsilon$ indicates the number of terms employed to evaluate \eqref{eq:pdf_snr}--\eqref{eq:cdf_snr}. From \eqref{eq:kappa:pdferror}, we follow similar steps as those leading to \eqref{eq:kappa:ub3} and, after some mathematical manipulations, we obtain
\begin{align}
    \mathcal{E}(\epsilon) < \ & \bigg(\frac{K}{\exp(\kappa)} \bigg)^{N \mu} \frac{8 N \Gamma (\mu +\epsilon) w^{N \mu+\epsilon-1+\zeta}}{7 \Gamma (\epsilon) \Gamma (N \mu +\epsilon+\zeta) \hat{w} ^{N \mu+\epsilon}} \nonumber \\
    & \times \tilde{K}^{\epsilon} \, _2F_2\bigg(1,\mu +\epsilon;\epsilon,N \mu+\epsilon+\zeta;\frac{\tilde{K} w}{ \hat{w}}\bigg). \label{eq:kappa:truncerror}
\end{align}

\subsection{Computational Complexity and Implementation}

\begin{figure}[t]
\centering
\includegraphics{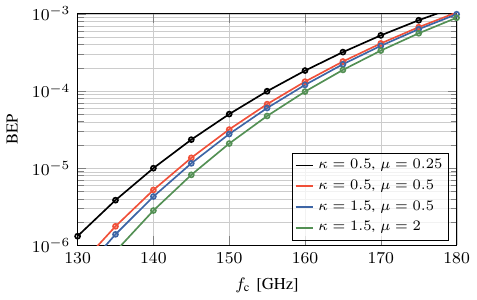}
\caption{\ac{PDF} of the sum of squared \ac{i.i.d.} \km{} \acp{RV}, with $N=64$, $\hat{w}=1$, and the following combinations of parameters: (i) $\kappa \rightarrow 0$, $\mu = 0.5$ (Nakagami-$m$); (ii) $\kappa = 1.5$, $\mu = 0.5$; (iii) $\kappa = 1.5$, $\mu = 1$ (Rice); and (iv) $\kappa = 1.5$, $\mu = 1.5$.}
\label{fig:pdf-kappa}
\end{figure}

As evinced by \eqref{eq:km}, the proposed framework uses recursion to numerically evaluate the derived expressions. A naive implementation would compute the same elements multiple times, which is highly inefficient when a large number of terms is involved. For example, by using the recursion tree method, the number of computations of the type of  \eqref{eq:km} amounts to $2^{\epsilon} - 1$. To reduce the computational complexity, we resort to memoization, a code optimization method that caches the value of every new recursive element, preventing the framework from computing it multiple times. With memoization, the number of computations of the type of \eqref{eq:km} reduces to $\epsilon (\epsilon + 1) - 1$, enabling fast evaluation even with a large number of terms. Fig.~\ref{fig:pdf-kappa} plots the analytical and simulated \acp{PDF}, with $N=64$, $\hat{w} =1$, and different combinations of fading parameters. The analytical \ac{PDF} is computed via \eqref{eq:pdf_snr} with truncation at $\epsilon = 250$ terms, and strikingly agrees with the simulations. Remarkably, the computation time for each curve in the plots peaks at about $0.5$~s and the maximum absolute error with respect to the exact solution in~\cite{Mil08} is in the order of $10^{-16}$.

\section{Performance Metrics}\label{sec:metrics}

In this section, considering the system model and the results for the \km{} distribution derived in Section~\ref{sec:newrep}, we obtain new expressions for the coverage probability and \ac{BEP} for coherent binary modulations. To evaluate the system's performance at high \ac{SNR}, we further derive the asymptotic expressions for these metrics. In~\cite{Tej25}, we additionally analyze the symbol error probability for $M$-ary phase-shift keying and quadrature amplitude modulation.

\subsection{Coverage Probability}

The coverage probability is defined as the probability that the instantaneous \ac{SNR} exceeds a given threshold. This can be readily formulated in terms of the \ac{CDF} of the instantaneous \ac{SNR}, i.e., $P_\textrm{cov} = 1 - F_{\gamma}(\gamma_\textrm{th})$, where $\gamma_\textrm{th}$ is the \ac{SNR} threshold. Hence, the coverage probability is readily available from \eqref{eq:cdf_snr}, where $\hat{w}$ is the scale parameter that can be used to model the path loss. To derive the asymptotic expression at high \ac{SNR}, as done in~\cite{Alm23b}, we use only the first term of the summation (corresponding to $m=0$) in \eqref{eq:cdf_snr}, resulting in
\begin{align} \label{eq:kappa_Pout_as}
    P_{\textrm{cov}} \sim 1 - \frac{1}{\Gamma (N \mu  +1)} \left(\frac{(\kappa +1) \mu \gamma_\textrm{th}}{\exp(\kappa) \hat{w}}\right)^{N \mu }.
\end{align}

\begin{figure*}
\setcounter{equation}{29}
\begin{align}
   \mathcal{E} (\epsilon) < \, _3F_2\left(N \mu+\epsilon+\frac{1}{2},\mu +\epsilon,1;N \mu+\epsilon+1,\epsilon;\frac{K}{g_{\textrm{b}} \hat{w} }\right) \frac{4 N\Gamma \left(\mu +\epsilon\right) \Gamma \big(N \mu +\epsilon+\frac{1}{2}\big)}{7 \sqrt{\pi} \Gamma \left(\epsilon\right) \Gamma \left(N \mu +\epsilon+1\right)} \bigg(\frac{K}{g_{\textrm{b}} \hat{w}}\bigg)^{\epsilon} \left(\frac{(\kappa +1) \mu }{\exp (\kappa ) g_{\textrm{b}} \hat{w}}\right)^{N \mu} \label{eq:kappa:bep-trunc}
\end{align}
\setcounter{equation}{24}
\hrulefill
\vspace{-4mm}
\end{figure*}

\subsection{\ac{BEP} for Coherent Binary Modulations}

\vspace{-0.5mm}

We now focus on the (uncoded) \ac{BEP} for coherent binary modulations. First, we obtain a new expression for the \ac{BEP} by resorting to the \ac{PDF} in \eqref{eq:pdf_snr}, along with its asymptotic expressions at high \ac{SNR}. Then, by deriving an upper bound on the \ac{BEP}, we determine the convergence condition for this expression, which proves to be quite restrictive. To guarantee the convergence for any choice of parameters, we thus propose an alternative expression for the \ac{BEP} based on the \ac{PDF} in \eqref{eq:kappa:pdf_snr2}. Lastly, we obtain an upper bound on the truncation error.

The \ac{BEP} for coherent binary modulations is defined as~\cite[Eq.~(9.3)]{Sim05} \vspace{-0.5mm}
\begin{align} \label{eq:pb}
    \textrm{BEP} = \frac{1}{2}\int_{0}^{\infty}\mathrm{erfc}(\sqrt{g_{\textrm{b}} w}) f_{W}(w) \diff w,
\end{align}
with $g_{\textrm{b}} = 1$ for coherent \ac{BPSK}, $g_{\textrm{b}} = 1/2$ for coherent orthogonal \ac{BFSK}, and $g_{\textrm{b}} = 0.715$ for coherent \ac{BFSK} with minimal correlation~\cite{Sim05}. In our case, $f_{W}(w)$ is replaced by the \acp{PDF} in \eqref{eq:pdf_snr} (or \eqref{eq:kappa:pdf_snr2}).

\smallskip

\textit{\textbf{Approach 1.}} By substituting  \eqref{eq:pdf_snr} into \eqref{eq:pb}, we obtain
\begin{align}
    \textrm{BEP} = \ & \frac{1}{2\sqrt{\pi}} \left(\frac{(\kappa +1) \mu}{\exp\left(\kappa\right)}\right)^{N \mu}\sum _{m=0}^{\infty }  \left(\frac{1}{g_{\textrm{b}} \hat{w} }\right)^{N \mu + m} \nonumber \\
    & \times \frac{ \Gamma \big(N \mu + m +\frac{1}{2}\big)k_m}{\Gamma (N \mu + m +1)} \label{eq:kappa:pb}.
\end{align}
To derive the asymptotic expression at high \ac{SNR}, we use only the first term of the summation (corresponding to $m=0$) in \eqref{eq:kappa:pb}, resulting in\footnote{Note that \eqref{eq:kappa:pb-asymp} can also be obtained as a special case of the expression provided in~\cite{Dix23}.} \vspace{-0.5mm}
\begin{align}\label{eq:kappa:pb-asymp}
    \textrm{BEP} \sim \ \frac{ \Gamma \big(N \mu  +\frac{1}{2}\big)}{2\sqrt{\pi} \Gamma \big(N \mu  +1\big)} \left(\frac{(\kappa +1) \mu}{\exp\left(\kappa\right) g_{\textrm{b}} \hat{w}}\right)^{N \mu}.
\end{align}
Now, following similar steps as in Section~\ref{sec:trunc}, we upper bound \eqref{eq:kappa:pb} as
\begin{align}
    \mathcal{C} < \ & 
    \frac{1 }{\hat{w} ^{N \mu}}  \Bigg(\frac{\Gamma\big(N\mu +\frac{1}{2}\big)}{\Gamma(N\mu + 1)} +\frac{8 N \Gamma (\mu +1) \Gamma \left(N \mu +\frac{3}{2}\right)}{7 \Gamma (N \mu + 2)} \nonumber \\
    & \times \frac{K}{g_{\textrm{b}} \hat{w} } \, _2F_1 \bigg(\mu +1;N \mu +\frac{3}{2};N \mu  +2;\frac{K}{g_{\textrm{b}}\hat{w} }\bigg)\Bigg). \label{eq:kappa:pb-converg}
\end{align}
Despite the strict convergence interval of the Gauss hypergeometric function $_2F_1$, common mathematical packages such as Mathematica and SciPy typically provide solutions with arguments outside the convergence interval through its analytical continuation. In our case, this feature could not be straightforwardly exploited for the summation in  \eqref{eq:kappa:pb}, restricting its convergence to choices of parameters satisfying $\big|K/ (g_{\text{b}} \hat{w})\big| < 1$. To circumvent this issue and guarantee the convergence for any choice of parameters, we propose an alternative expression in the following.

\textit{\textbf{Approach 2.}} By substituting \eqref{eq:kappa:pdf_snr2} into \eqref{eq:pb}, we obtain (cf. \eqref{eq:kappa:pb})
\begin{align}
    \textrm{BEP} = \ & \sum _{m=0}^{\infty } \frac{ \Gamma \big(N \mu \! + \! m \! + \!\frac{1}{2}\big) \big(\frac{\tilde{K}}{g_{\textrm{b}} \hat{w}}\big)^{N \mu + m} \tilde{k}_m}{ 2\sqrt{\pi} \exp(N\kappa\mu) \Gamma (N \mu \!+ \!m \! + \! 1)} \, _2F_1\bigg(N \mu \! + \! m, \nonumber \\
    & N \mu + m+\frac{1}{2};N \mu + m+1;\frac{ -\tilde{K} }{g_{\textrm{b}}  \hat{w} }\bigg) \label{eq:kappa:pbfinal2}.
\end{align}
Remarkably, \eqref{eq:kappa:pbfinal2} allows to evaluate the \ac{BEP} for any choice of parameters at the cost of slightly increased mathematical and computational complexity due to the presence of the Gauss hypergeometric function $_2F_1$. However, the computational complexity associated with this function can be drastically reduced by using its recurrence property~\cite[Eq. (15.5.E19)]{NIST:DLMF}.

\smallskip

Lastly, by replicating the steps in Section~\ref{sec:trunc}, we obtain an upper bound on the truncation error for the \ac{BEP} in \eqref{eq:kappa:bep-trunc} at the top of the page, which applies to both approaches.

\section{Performance Evaluation} \label{sec:num}

In this section, we build on the proposed framework for the \km{} model to analyze the performance of an uplink system operating in the sub-THz band. As in the system model described in Section~\ref{sec:newrep}, we consider a single-antenna user served by a multi-antenna \ac{BS} employing \ac{MRC}, with perfect and imperfect \ac{CSI}. We compare the analytical results (obtained using Mathematica on a standard off-the-shelf laptop computer) with Monte Carlo simulations. In all the plots, the lines indicate the analytical results obtained with the proposed framework and the markers represent the simulation results, which are in agreement with the analytical results.

We assume a log-distance path loss model according to which we have $\hat{w} = \frac{P_{\textrm{t}}}{\sigma^{2}} \varphi d^{-\beta}$, where $\varphi$ represents the frequency-dependent path loss factor referenced at 1~m, $d$ denotes the distance between the user and the BS, and $\beta = 2$ models the \ac{LoS}-dominated propagation. The path loss factor is defined as $\varphi = \big( c/(4\pi f_\textrm{c}) \big)^{2}$, where $c$ represents the speed of light and $f_\textrm{c}$ denotes the carrier frequency. The noise power is calculated as $\sigma^2 = -174 + 10\log_{10} \Omega+ \nu$ (measured in dBm), where $\Omega$ denotes the transmission bandwidth and $\nu = 6$~dB is the noise figure of the \ac{BS}. The bandwidth is set to $0.5 \%$ of the carrier frequency, i.e., $\Omega = 0.005 f_{\textrm{c}}$. Unless otherwise stated, we fix $P_{\textrm{t}} = 23$~dBm, $f_{\textrm{c}} = 140$~GHz, $d=200$~m, and assume perfect \ac{CSI} at the \ac{BS} (i.e., $\alpha = 0$); furthermore we assume $\kappa = 1.5$ and $\mu = 0.5$ for the fading parameters.\footnote{The fading parameters depend on the specific propagation conditions and fitting them to realistic scenarios is beyond the scope of this paper. Instead, our focus is to demonstrate the efficiency of the proposed framework and evaluate the system's performance across different choices of parameters.}

\begin{figure}[t]
\centering
\includegraphics{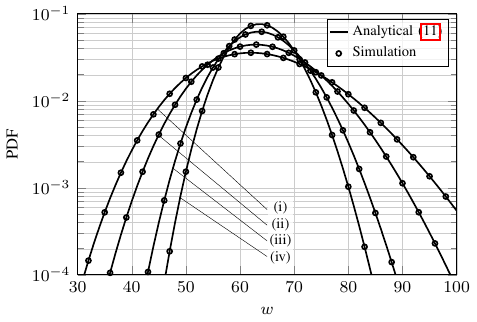}
\caption{Uplink coverage probability versus distance, with $\kappa = 1.5$, $\mu = 0.5$, $f_\textrm{c} = 140$~GHz, $P_{\textrm{t}} = 23$~dBm, and $\alpha=0$.}
\label{fig:kappa-coverage2}
\end{figure}

\begin{figure}[t]
    \centering
    \includegraphics{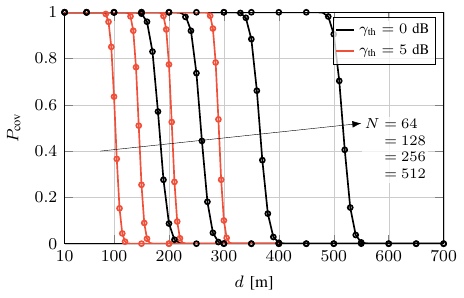}
    \caption{Uplink \ac{BEP} for coherent \ac{BPSK} versus transmit power, with $\kappa = 1.5$, $\mu = 0.5$, $f_\textrm{c} = 140$~GHz, and $d = 200$~m.}
    \label{fig:kappa-bep-pt}
\end{figure}

Fig.~\ref{fig:kappa-coverage2} plots the coverage probability versus the distance between the user and the \ac{BS}, with $\gamma_\textrm{th} \in \{0, 5\}$~dB.
Adopting $N=512$ allows to compensate for the strong path loss at $140$~GHz and achieve perfect coverage up to approximately $250$~m and $450$~m for $\gamma_{\textrm{th}} =5$~dB and $\gamma_{\textrm{th}} =0$~dB, respectively. Note that this extensive coverage is primarily enabled by the truly massive number of antennas combined with the \ac{LoS}-dominated propagation. The analytical results are computed with $500$ terms, remarkably taking less than $5$~s to generate the whole plot. For comparison, the expression in~\cite{Mil08} provides the same output in about $105$~s.

Fig.~\ref{fig:kappa-bep-pt} illustrates the \ac{BEP} for coherent \ac{BPSK} versus the transmit power along with the asymptotic \ac{BEP} at high \ac{SNR} (dashed lines), with $\alpha\in \{0,0.4\}$. Note that the asymptotic \ac{BEP} is only depicted for $N=64$ to improve the readability of the plot. For $P_\textrm{t} = 23$~dBm, doubling the number of antennas from $N = 256$ to $N = 512$ results in a $36 \times$ reduction in the \ac{BEP} with $\alpha = 0$, i.e., from $5 \times 10^{-3}$ to $1.4 \times 10^{-4}$. Furthermore, the asymptotic slopes are mainly dictated by the number of antennas and the number of multipath clusters. The analytical results are computed with no more than $500$ terms and the entire plot is generated in less than $3$~s.

\begin{figure}[t]
    \centering
    \includegraphics{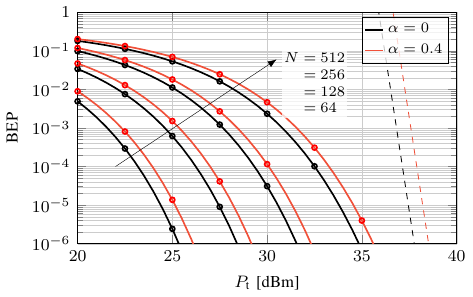}
    \caption{Uplink \ac{BEP} for coherent \ac{BPSK} versus carrier frequency, with $\kappa = 1.5$, $\mu = 0.5$, $d = 200$~m, and $P_{\textrm{t}} = 23$~dBm.}
    \label{fig:kappa-bep-freq}
\end{figure}

Fig.~\ref{fig:kappa-bep-freq} depicts the \ac{BEP} for coherent \ac{BPSK} versus the carrier frequency, again with $\alpha\in\{0, 0.4\}$. Fixing a target \ac{BEP} of $10^{-3}$, increasing the number of antennas from $N=128$ to $N=512$ allows the carrier frequency (and thus the bandwidth) to be increased by about 60\%.

\begin{figure}[t]
    \centering
    \includegraphics{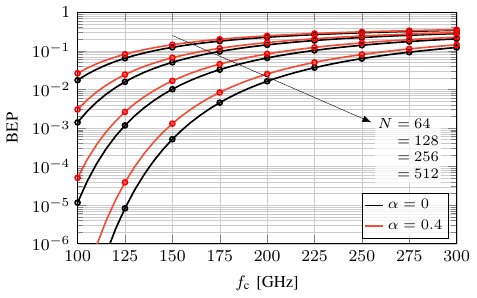}
    \caption{Uplink \ac{BEP} for coherent \ac{BPSK} versus carrier frequency, with $N=128$, $d = 80$~m, $P_{\textrm{t}} = 23$~dBm, and $\alpha =0$.}
    \label{fig:kappa-bep-fading}
\end{figure}

Fig.~\ref{fig:kappa-bep-fading} shows the \ac{BEP} for coherent \ac{BPSK} versus the carrier frequency, with $N = 128$. We consider four combinations of fading parameters mimicking fading conditions ranging from very severe to favorable: (i) $\kappa=0.5$, $\mu=0.25$ (black line); (ii) $\kappa=0.5$, $\mu=0.5$ (red line); (iii) $\kappa=1.5$, $\mu=0.5$ (blue line); and (iv) $\kappa=1.5$, $\mu=2$ (green line). In (i), the \ac{BEP} is undermined by the small value of $\mu$ (i.e., the number of multipath clusters). In (ii), by slightly increasing $\mu$, the \ac{BEP} significantly improves, which allows to push the carrier frequency up by at least $2$~GHz. In (iii), a $3 \times$ increase in $\kappa$ (i.e., the power ratio between the \ac{LoS} and \ac{NLoS} components) produces only a timid decrease in the \ac{BEP} since the array gain weakly depends on $\kappa$. Lastly, in (vi), a bigger increase in $\mu$ yields a considerable \ac{BEP} improvement, further raising the carrier frequency by at least $1$~GHz. Hence, even though we consider \ac{LoS}-dominated propagation, the \ac{NLoS} components still have a noticeable impact on the system's performance.

\section{Conclusions} \label{sec:final}

In this paper, we developed a new exact representation of the sum of squared \ac{i.i.d.} \km{} \acp{RV}. The proposed analytical framework is remarkably tractable and computationally efficient, and thus can be conveniently employed to analyze system with massive antenna arrays. We derived novel expressions for the \ac{PDF} and \ac{CDF}, we analyzed their convergence and truncation error, and we discussed the computational complexity and implementation aspects. Furthermore, we derived expressions for the  coverage probability and \ac{BEP} for coherent binary modulations. Lastly, we evaluated the performance of an uplink sub-THz system where a single-antenna user is served by a \ac{BS} employing \ac{MRC}. The results revealed that the number of antennas and the number of multipath clusters have the greatest impact on the system's performance. Future work will explore extensions to the multi-user scenario.

\bibliographystyle{IEEEtran}
\bibliography{refs_abbr,refs}

\end{document}